\documentclass[%
reprint,
amsmath,
amssymb,
aps,
]{revtex4-2}
\usepackage{xcolor}
\usepackage[caption=false]{subfig}
\usepackage{graphicx}% Include figure files
\usepackage{dcolumn}% Align table columns on decimal point
\usepackage{bm}% bold math
\definecolor{violet}{rgb}{0.592,0.474,0.957}
\definecolor{red}{rgb}{0.957,0.427,0.490}
\usepackage[colorlinks=true,urlcolor=violet,linkcolor=red,citecolor=violet]{hyperref}
\usepackage{mleftright}

\usepackage{xr}
\makeatletter
\newcommand*{\addFileDependency}[1]{% argument=file name and extension
  \typeout{(#1)}
  \@addtofilelist{#1}
  \IfFileExists{#1}{}{\typeout{No file #1.}}
}
\makeatother

\newcommand*{\myexternaldocument}[1]{%
    \externaldocument{#1}%
    \addFileDependency{#1.tex}%
    \addFileDependency{#1.aux}%
}

\myexternaldocument{SupplementalMaterials}

\begin{document}

\preprint{APS/123-QED}

\title{Spontaneous Demixing of Binary Colloidal Flocks}% Force line breaks with \\

\author{Samadarshi Maity}
 \affiliation{Huygens-Kamerlingh Onnes Laboratory, Universiteit Leiden,
PO Box 9504, 2300 RA Leiden, the Netherlands
}
\author{Alexandre Morin*}%

\affiliation{Huygens-Kamerlingh Onnes Laboratory, Universiteit Leiden,
PO Box 9504, 2300 RA Leiden, the Netherlands
}



%v Add the date her 
\date{\today}

\begin{abstract}
Population heterogeneity is ubiquitous among active living systems, but little is known about its role in determining their spatial organization and large-scale dynamics.
Combining evidence from synthetic active fluids assembled from self-propelled colloidal particles along with theoretical predictions at the continuum scale, we demonstrate the spontaneous demixing of binary polar liquids within circular confinement.
Our analysis reveals how both active speed heterogeneity and non-reciprocal repulsive interactions lead to self-sorting behavior.
By establishing general principles for the self-organization of binary polar liquids, our findings highlight the specificity of multi-component active systems.
\end{abstract}

\maketitle

\paragraph*{Introduction.—}

In the savanna, zebras often herd together with giraffes to benefit from their vigilance~\cite{schmitt2016zebra}.
The formation of such heterogeneous group is not the exception.
Often in living systems, diverse groups gather and move collectively:
from human crowds, to bird flocks, to bacterial colonies~\cite{elias2012multi,ben2016multispecies,ward2018cohesion,peled2021heterogeneous,ariel2022variability}.
In stark contrast, our current account of motile active matter is mostly restricted to single-component systems~\cite{marchetti2013hydrodynamics,gompper20202020,bowick2022symmetry}:
focusing on homogeneous populations, physicists have captured the emergence of a variety of active phases, such as the flocking in polar liquids~\cite{vicsek1995novel,toner2005hydrodynamics,bricard2013emergence}, the turbulence in active nematics~\cite{sanchez2012spontaneous,giomi2015geometry,doostmohammadi2018active}, and the motility-induced phase separation of active Brownian particles~\cite{theurkauff2012dynamic,buttinoni2013dynamical,palacci2013living,cates2015motility}.
Elucidating the impact of population heterogeneity on these collective behaviours represents a formidable yet necessary challenge. 
It calls for thorough experimental investigations~\cite{curatolo2020cooperative} to take further strides following recent numerical and theoretical insights~\cite{kolb2020active,pattanayak2020speed,adhikary2022pattern,zuo2020dynamic,ariel2015order,pessot2018binary,menzel2012collective,kolb2020statistical,chatterjee2023flocking,dinelli2022nonreciprocity,palmer2022hydrodynamic}:
Are active phases robust to such heterogeneities? Do they benefit from them? Do novel phases emerge?
In this Letter, we combine experiments and theory to investigate the flocking behavior of binary polar liquids and unveil an active phase whose components spontaneously demix.

\begin{figure}[b]
\centering
\subfloat{\includegraphics[width=\columnwidth]{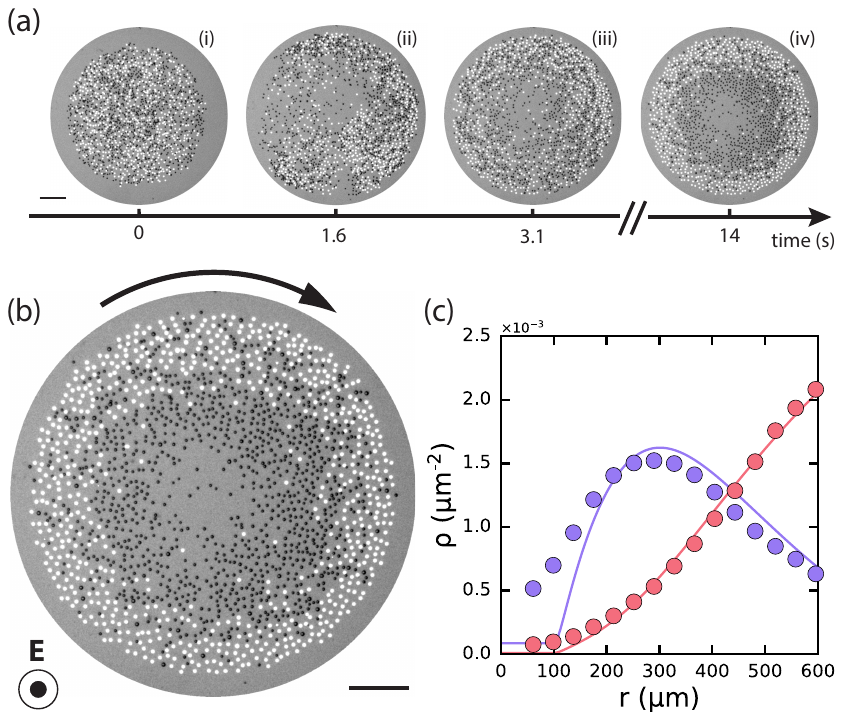}}
\caption{Spontaneous demixing in binary colloidal flocks. (a) A bidispersed mixtures of colloidal rollers, activated at $t=0$, forms a flock and eventually demixes. (b) The binary colloidal flock moves clockwise at steady-state. It is assembled from $N_\mu = 690$ $10\,\rm \mu m$-diameter fluorescent particles (bright) and $N_\nu = 1050$ $7\,\rm \mu m$-diameter non-fluorescent particles (dark) and confined in a circular well of $600\,\rm \mu m$-diameter. Colloidal particles are activated through the Quincke mechanism by application of the DC electric field $\mathbf{E}$.  (c) Radial density profiles of both species for a 52:48 binary mixture with $N_\mu =1360$ (red) and  $N_\nu = 1260$ (violet). The density of $10\,\rm \mu m$ particles increases from the core to the edge, while the density of $7\,\rm \mu m$ particles display non-monotonous behavior. Scale bars: $200\,\rm \mu m$.
}
\label{Fig:SpontaneousDemixing}
\end{figure}

Heterogeneity of active components can influence different aspects of their dynamics.
It can affect their individual motion, such as their self-propulsion speed~\cite{kolb2020active,pattanayak2020speed,adhikary2022pattern,zuo2020dynamic} or their intrinsic fluctuations~\cite{ariel2015order,pigolotti2014selective,ilker2020phase}, or it can alter their interactions. 
In particular, the out-of-equilibrium nature of active systems allows for non-reciprocal interactions resulting in asymmetric interactions between species with major consequences on their self-organization~\cite{ivlev2015statistical,meredith2020predator,dinelli2022nonreciprocity,soto2014self,agudo2019active,curatolo2020cooperative,fruchart2021non,you2020nonreciprocity,chatterjee2023flocking}.
Here, we consider mixtures of active colloids that are bi-dispersed in size and demonstrate the ensuing heterogeneities of both their individual activity and interactions rules.
By assembling colloidal flocks~\cite{bricard2013emergence} from such binary mixtures, we uncover their spontaneous demixing, see Fig.~\ref{Fig:SpontaneousDemixing}(b).
We explain this behavior by developing a hydrodynamic theory for binary polar liquids which captures our experimental observations without any free parameter, see Fig.~\ref{Fig:SpontaneousDemixing}(c).
Our analysis eventually unveils two generic mechanisms for the spontaneous demixing in polar liquid vortices based on self-propulsion speed differences and non-reciprocal repulsive interactions.

\paragraph*{Demixing in Binary Colloidal Flocks.—}

Figure~\ref{Fig:SpontaneousDemixing}(a) illustrates a typical experiment where binary colloidal flocks self-assemble from colloidal rollers confined within circular microfluidic wells~\cite{bricard2013emergence,bricard2015emergent}.
The initially homogeneously-distributed mixture (i) quickly self-organizes into a dynamical vortex (ii)-(iii) upon activation of the particles at $t=0$.
A few seconds later, a striking feature appears within the binary flock (iv): its constituents have spontaneously demixed!
This spatial organization, which corresponds to the system steady-state, consists in the accumulation of the larger particles towards the edge, while the smaller ones populate the core.
As shown in the Supplemental Materials (SM), this partial demixing persists through variations in the confinement size, the mixture ratio, and the particle material.

To understand the origin of this behavior, let us first delve into the activation mechanism and the emergence of the polar liquid vortex.
We focus here on some key elements and provide additional details in the SM.
To investigate the influence of heterogeneity on their collective dynamics, binary flocks are obtained by combining polystyrene colloidal particles of two different sizes: $a_\mu = 10\,\rm \mu m$- and $a_\nu = 7\,\rm \mu m$ in diameter\textcolor{black}{, of uniformity below $5\,\%$, see SM Section~\ref{Sec:Experiments}.} The subscripts $\mu$ and $\nu$, each refer to the respective species throughout this Letter.
The activation of the particles is achieved by means of the Quincke electro-hydrodynamic instability which causes the particles to rotate at a constant rate independent of their size~\cite{quincke1896ueber,das2013electrohydrodynamic}.
This steady rotation is converted into translation {\it via} friction with the substrate resulting in their self-propulsion at fixed speed~\cite{bricard2013emergence}.
This activation mechanism gives rise to the so-called {\em colloidal rollers}, that perform persistent random walks when isolated~\cite{bricard2013emergence}.
Importantly for the study of binary flocks, differences in particle size yield differences in the self-propulsion speed $v$ and the rotational diffusivity $D$ of each species.
Under typical experimental conditions, we measure $v_\mu = 0.9\,\rm mm\,s^{-1}$, $D_\mu = 1.9\,\rm s^{-1}$ and $v_\nu = 0.7\,\rm mm\,s^{-1}$, $D_\nu = 3.4\,\rm s^{-1}$, see SM Section~\ref{Sec:Experiments}.
We note that $a_\nu/a_\mu \neq v_\nu/v_\mu$, which reveals subtle effects of the substrate on their individual dynamics~\cite{pradillo2019quincke,zhang2021quincke}.

At high density, an initially homogeneous, binary mixture eventually forms a polar liquid vortex  with radially increasing density, similar to single-species flocks~\cite{bricard2015emergent}.
Figure~\ref{Fig:SpontaneousDemixing}(b) shows such a colloidal vortex for a 40:60 mixture consisting of $N_\mu = 690$ large and $N_\nu = 1050$ small particles, see also the Supplementary Movie.
In populations of colloidal rollers, collective motion can be traced back to alignment interactions between rollers mediated by the solvent (see~\cite{bricard2013emergence} and below).
Upon increase in the roller density, alignment interactions eventually overcome rotational diffusion and the population undergoes a Vicsek-like transition~\cite{vicsek1995novel}.
In circular confinement, collective motion takes the form of a vortex that rotates with equal probability in the clockwise or counter-clockwise direction breaking the initial symmetry of the system~\cite{bricard2015emergent,seyed2016vortex,chardac2021emergence,zhang2022polar}.
\textcolor{black}{
In this context, the formation of a polar liquid vortex by the binary mixture echoes the single-component case:
50:50 mixtures form a gas at low density and a polar vortex at high density, see SM Sec.~\ref{Sec:Experiments}.
In this Letter, we focus on densities above the flocking threshold, and particularly on the spontaneous demixing taking place within the binary flocks.
}
%Exclusive to the binary flock, however, is the spontaneous demixing taking place within the mixture.
%

What is the origin of this spontaneous demixing?
Not only size heterogeneity causes differences in the intrinsic properties of active particles, $v$ and $D$, but it also results in a unique set of pairwise interactions.
In particular, repulsive interactions can be expected to play a major role: in single-component polar liquid vortices, self-propulsion and repulsion compete and their balance sets the radial increase of the density profile~\cite{bricard2015emergent}.
In binary flocks, distinct density profiles suggest that each species achieves this balance in its own way.
Unraveling what determines the spatial structure of binary flocks therefore requires establishing the interactions rules intra- and inter-species to identify the consequence of size heterogeneity.
Beyond colloidal flocks, sourcing what factor drives the spatial structuring of binary flocks would unveil universal features yielding spontaneous demixing.

\paragraph*{Non-reciprocal interactions.—}

\begin{figure}[t]
\subfloat{\includegraphics[width=0.75\columnwidth]{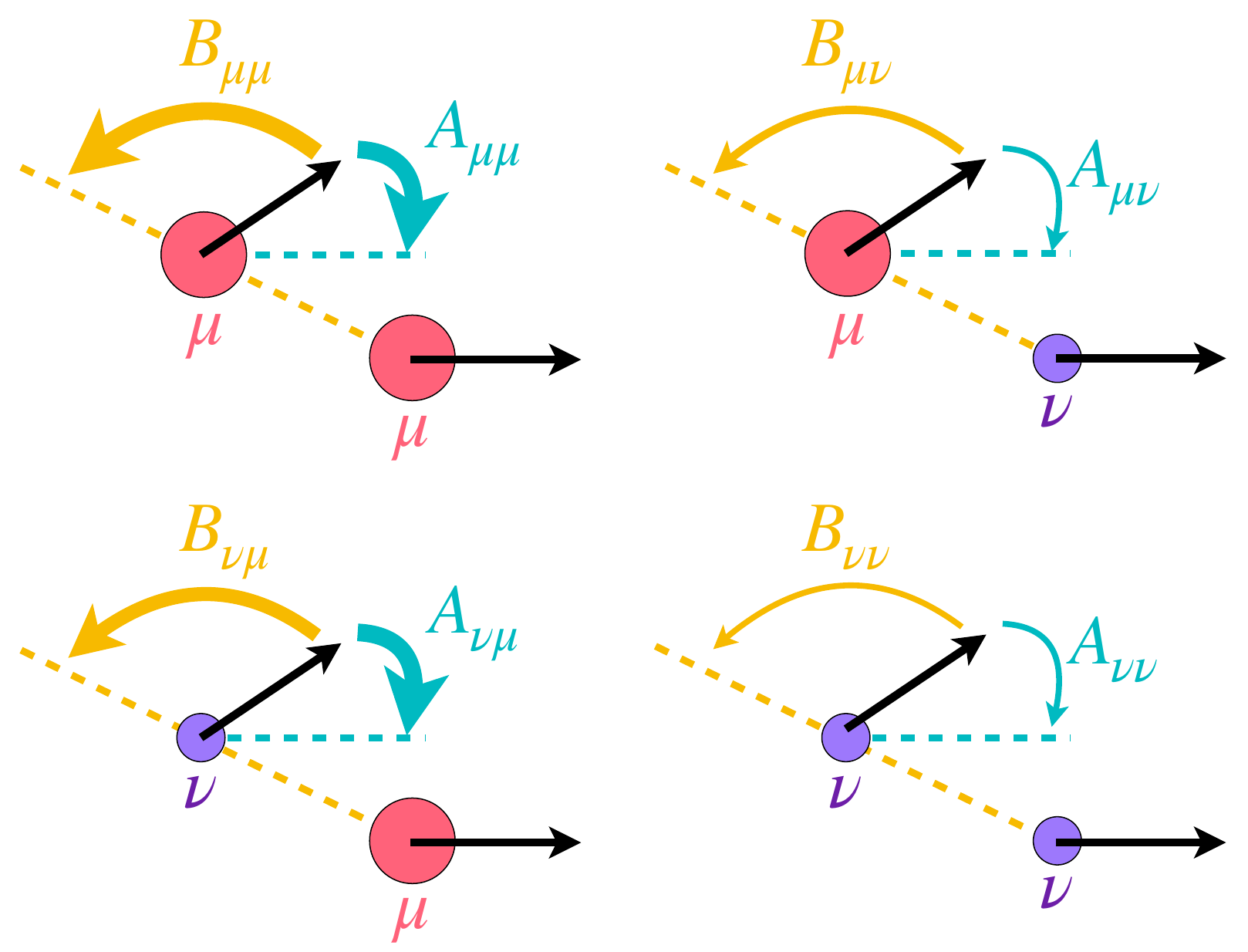}}%
\caption{Non-reciprocal interactions between colloidal rollers. Pairwise interactions consist in two torques acting on the direction of self-propulsion (black arrows). The alignment torque $A$ originates from hydrodynamic coupling and favors self-propulsion in a common direction. The repulsive torque $B$ originates from electrical coupling and favors self-propulsion away from the center-to-center segment. The thickness of the colored arrows materializes the non-reciprocity of the interactions, $A_{\mu\mu} = A_{\nu\mu} > A_{\mu\nu} = A_{\nu\nu}$ and $B_{\mu\mu} > B_{\nu\mu} > B_{\mu\nu} > B_{\nu\nu}$, see Eqs.~\eqref{Eq:NonReciprocalAli} and~\eqref{Eq:NonReciprocalRep}.}%
\label{Fig:NonReciprocity}
\end{figure}

Interacting colloidal rollers are well described by a generic active XY model with alignment and repulsive couplings~\cite{bricard2013emergence,caussin2014tailoring}, as illustrated in Fig.~\ref{Fig:NonReciprocity}.
These interactions take their roots in the propulsion mechanism of Quincke rollers and originate respectively from hydrodynamic and electrical interactions.
Keeping the dominating contributions only, binary interactions lead to the relaxation of the particle velocity orientation $\mathbf{\hat{v}_{i}} = \left( \cos\theta_i,\, \sin\theta_i \right)$ in a characteristic time $\tau$ in a potential that is a function of the orientations of both particles $\mathbf{\hat{v}_{i}}$ and $\mathbf{\hat{v}_{j}}$ and their interparticle distance $\mathbf{r_i} - \mathbf{r_j} = r\mathbf{\hat{r}}$:
\begin{equation}
    \frac{{\rm d}\theta_i}{{\rm d} t} = \frac{1}{\tau}\frac{\partial}{\partial \theta_i}  \left( A_{ij}(r)\mathbf{\hat{v}_i}\cdot \mathbf{\hat{v}_j}  +   B_{ij}(r)\mathbf{\hat{v}_i}\cdot \mathbf{\hat{r}}  \right),
     \label{Eq:Interactions}
\end{equation}
where $A_{ij}(r)$ and $B_{ij}(r)$ are the alignment and repulsion strengths caused by particle $j$ on particle $i$, respectively.
Alignment and repulsion need not be reciprocal when considering interactions between heterospecifics as a consequence of the non-equilibrium nature of the system.
Indeed, their expressions, derived from first principles in the SM Section~\ref{Sec:QuinckeAndCo}, are asymmetrical under $i \leftrightarrow j$ inversion which reveals such non-reciprocity:
\begin{align}
    &A_{ij}(r) = \mathcal{A} \frac{a_j^3}{r^3} \Theta(r), \label{Eq:NonReciprocalAli}\\
    &B_{ij}(r) = \mathcal{B}  \frac{a_j^3 a_i}{r^4}  \Theta(r). \label{Eq:NonReciprocalRep}
\end{align}
Here, $\mathcal{A}$ and $\mathcal{B}$ are constants and $\Theta(r)$ is a screening function indicating that interactions are short-ranged.
While we keep the details of the derivation in the SM, Eqs.~\eqref{Eq:NonReciprocalAli} and~\eqref{Eq:NonReciprocalRep} are worth a few comments.
First, the hydrodynamic interaction shows no dependency on the size of particle $i$.
This feature originates from the fact that colloidal rollers respond identically to the shear of hydrodynamic flows~\cite{morin2018flowing} irrespective of their size.
Second, the repulsive torques originating from electrical interactions are also non-reciprocal.
This feature may seem even more surprising than for viscous hydrodynamic interactions. 
In fact, colloidal rollers are constantly powered by the application of a DC electric field which builds dipolar charge distributions around the particles.
Continuous charge transfer with the surrounding liquid breaks force parity.
Finally, within binary mixtures, the above pairwise interactions imply that only two distinct coefficients describe alignment interactions while four different coefficients are necessary to capture repulsive interactions.
Figure~\ref{Fig:NonReciprocity} summarizes the interactions between conspecific and heterospecific colloidal rollers.

Overall, size differences of colloidal rollers give rise to rich microscopic dynamics where particles self-propel at different speeds and interact according to complex rules.
Therefore, spontaneous demixing of binary flocks result {\it a priori} from both intrinsic characteristics and interaction parameters.

\paragraph*{Hydrodynamics of binary polar liquids.—}

In order to capture the structuring of binary flocks at the macroscopic level, we employ a hydrodynamic description of binary polar liquids.
By coarse-graining the microscopic equations of motions of particles of species $\mu$ and $\nu$, we establish the evolution equations for the density fields $\rho_\mu$, $\rho_\nu$, and the polarization fields $\mathbf{\Pi}_\mu$, $\mathbf{\Pi}_\nu$, akin to the Toner-Tu equations for single-component polar liquids~\cite{toner1995long}, see SM Section~\ref{Sec:HydroTheory}.
The density and momentum dynamics of species $\mu$ read, respectively:
\begin{equation}
\partial_t \rho_\mu + \nabla\cdot \left( \rho_\mu\mathbf{\Pi}_\mu \right) = 0
\label{Eq:massBinary}
\end{equation}
\begin{equation} 
    \begin{split}
    \partial_t(\rho_\mu \boldsymbol{\Pi}_\mu &) + v_\mu \nabla.\left(\rho_\mu \textbf{Q}_\mu +\frac{\rho_\mu}{2}\textbf{I}\right) = 
     -D_\mu \rho_\mu\boldsymbol{\Pi}_\mu +\\& \rho_\mu(\textbf{I}-2\textbf{Q}_\mu).\left[\alpha_{\mu\mu}\rho_\mu\boldsymbol{\Pi}_\mu 
     + \alpha_{\mu\nu}\rho_\nu  \boldsymbol{\Pi}_\nu \right]
     \\& - \rho_\mu(\textbf{I}-2\textbf{Q}_\mu).\left[\beta_{\mu\mu}\nabla\rho_\mu + \beta_{\mu\nu}\nabla\rho_\nu\right],
    \end{split}
    \label{Eq:momentumBinary}
\end{equation}
and the ones of species $\nu$ follows from $\mu \leftrightarrow \nu$ inversion.
In Eq.~\eqref{Eq:momentumBinary}, $\mathbf{Q}$ is the nematic order parameter.
The coefficients $\alpha$ and $\beta$  express alignment and repulsion couplings at the hydrodynamic level and are obtained from their microscopic counterparts.
Interactions between conspecifics correspond to the $\alpha_{\mu\mu}$ and $\beta_{\mu\mu}$ terms while interactions between heterospecifics correspond to the $\alpha_{\mu\nu}$ and $\beta_{\mu\nu}$ ones.
We performed several experiments dedicated to measuring these coefficients, which we report in the SM Section~\ref{Sec:CoefficientsMeasurements}.
Our experiments corroborate that the alignment coefficients contain two distinct values, reflecting the theoretical description Eq.~\eqref{Eq:NonReciprocalAli}.
Experimentally, we could not differentiate $\beta_{\mu\mu}$ and $\beta_{\nu\mu}$, nor $\beta_{\nu\nu}$ and $\beta_{\mu\nu}$.
As a consequence, the alignment and repulsion coefficients read:

\begin{equation}
\begin{gathered}
\underline{\underline{\mathbf{\alpha}}} = \alpha_{\mu \mu}
\begin{pmatrix}
1 & 0.53 \\
1 & 0.53 
\end{pmatrix},\;\;\;\;\;\;\;\;
\underline{\underline{\mathbf{\beta}}} = \beta_{\mu \mu}
\begin{pmatrix}
1 & 0.58 \\
1 & 0.58 
\end{pmatrix},
\end{gathered}
\label{Eq:HydroCoefficients}
\end{equation}
where $\alpha_{\mu\mu} = 5.5\times10^4\,\rm \mu m^2/s$ and $\beta_{\mu\mu} = 2.7\times10^{5}\,\rm \mu m^3/s$.

To describe flocks confined within circular boundaries, we now look for axisymmetric stationary solutions by projecting the momentum equations onto the azimuthal $\mathbf{\hat{e}_\varphi}$ and radial $\mathbf{\hat{e}_r}$ directions.
The projections on the azimuthal direction highlight that alignment coupling and rotational noise compete to set the local polarizations (see SM Section~\ref{Sec:HydroTheory}).
It turns out that the two species show only minute differences in their polarization profiles, see SM Fig.~\ref{Fig:PolarizationProfiles}.
This allows to make an important simplification and consider $\mathbf{\Pi}_\mu(r) = \mathbf{\Pi}_\nu(r)$, $D_\mu = D_\nu$, and $\mathbf{Q}_\mu(r) = \mathbf{Q}_\nu(r)$ which follows from the closure relation $\mathbf{Q}(r) = \Pi^2(r)/2 \left(\mathbf{\hat{e}_\varphi}\mathbf{\hat{e}_\varphi} - \mathbf{\hat{e}_r}\mathbf{\hat{e}_r} \right)$.

The projections on the radial direction encode spontaneous demixing.
As anticipated, we observe that the density profiles are set by the competition between repulsion and an effective centrifuge force originating from the particle self-propulsion within the vortex, in line with \cite{bricard2015emergent}.
\begin{figure}[b]
    \includegraphics[width=0.5\columnwidth]{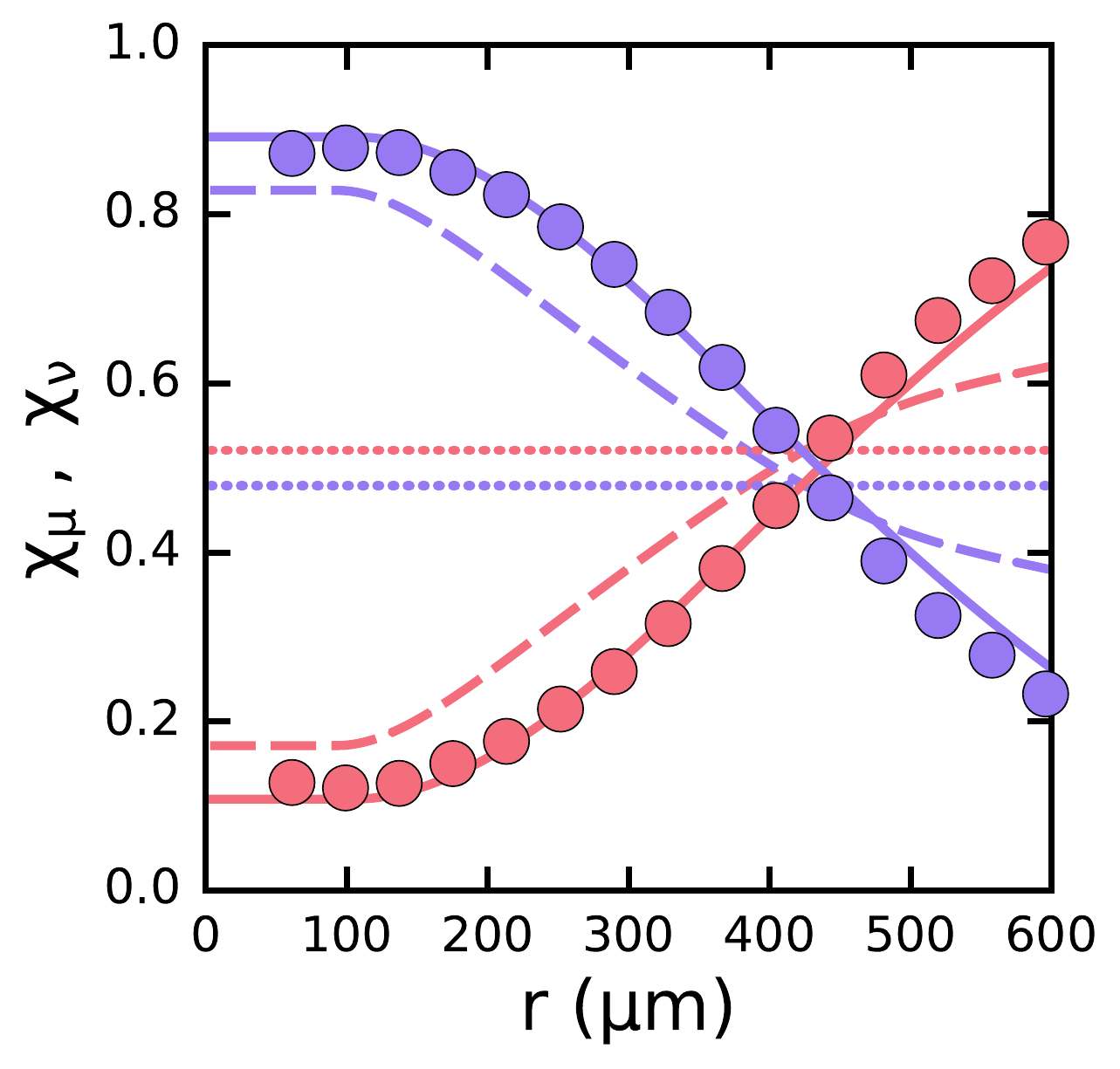}%
    \caption{Radial profiles of the density fractions $\chi_\mu$ (red) and $\chi_\nu$ (violet). The hydrodynamic theory (solid line) perfectly captures the monotonous radial evolutions of the experimental (markers) small (violet) and large (red) particle density fractions. The dashed line and the dotted lines indicate the theoretical predictions for reciprocal heterospecific interactions and for common self-propulsion speed, respectively.}%
\label{Fig:DensityProfilesFractions}
\end{figure}
This is most easily seen in the case where non-reciprocity is neglected.
In this case, the alignment and repulsion matrices reduce to single coefficients $\alpha$ and $\beta$, and the total density field $\rho_{\mu + \nu}(r) = \rho_\mu(r) + \rho_\nu(r)$ satisfies:
\begin{equation}
    \left( \frac{\Pi^2}{1+\Pi^2} \right) \frac{\overline{v(r)}}{r} = \beta \frac{\partial \rho_{\mu + \nu}}{\partial r},
    \label{Eq:Radial}
\end{equation}
where $\overline{v(r)}$ is the local net activity~\cite{kolb2020active}, defined here as $\overline{v} = \left( \chi_\mu/v_\mu + \chi_\nu/v_\nu \right)^{-1}$, where we introduce the local density fractions $\chi_\mu(r) = \rho_\mu/\rho_{\mu + \nu}$ and $\chi_\nu(r) = \rho_\nu/\rho_{\mu + \nu}$.
The left-hand side being always positive, Eq.~\eqref{Eq:Radial} predicts the monotonous radial increase of the total density\textcolor{black}{, as soon as the system is in its ordered state ($\Pi > 0$)}.
The demixing of the binary flock is in turn captured by the evolution of the density fractions, which relate to the total density variation {\it via}:
\begin{equation}
\frac{1}{\chi_\mu (1-\chi_\mu)} \frac{\partial \chi_\mu}{\partial r} = 
    K \left( \frac{v_\mu -v_\nu }{v_\mu v_\nu}\right) \rho_{\mu + \nu}  \frac{\partial \rho_{\mu + \nu}}{\partial r},
        \label{Eq:RadialFraction}
\end{equation}
where $K= 2 \alpha\beta \left( 1 + \Pi^2 \right)/D$ is always positive.
Equation~\ref{Eq:RadialFraction} predicts that the density fractions evolve monotonously according to the self-propulsion speed difference between the two species, in agreement with our experimental observations.
Figure~\ref{Fig:DensityProfilesFractions} shows the experimental profiles of $\chi_\mu$ and $\chi_\nu$ together with the theoretical predictions obtained by solving numerically Eqs.~\eqref{Eq:Radial} and~\eqref{Eq:RadialFraction} with no fitting parameter (dashed line): the fraction of the large, faster, species radially increases, and the fraction of the small, slower, species decreases.

This good qualitative agreement underscores the essential role of speed differences in the demixing of binary colloidal flocks.
To capture even more finely the structure of the binary flock, however, the non-reciprocity of interactions must be taken into account.
Factoring the non-reciprocities of Eq.~\eqref{Eq:HydroCoefficients} in the theory yields expressions ressembling Eqs.~\eqref{Eq:Radial} and~\eqref{Eq:RadialFraction}, which can be similarly integrated, see SM Section~\ref{Sec:HydroTheory}.
Doing so eventually leads to an excellent match between experiments and theory, as shown in Fig.~\ref{Fig:DensityProfilesFractions} (solid lines).
We stress that this quantitative agreement relies solely on hydrodynamic coefficients inferred experimentally, and is obtained without any fitting parameter.
Altogether, the hydrodynamic description reveals that active speed differences and non-reciprocal interactions are the necessary and sufficient ingredients to capture the spatial structuring of binary colloidal flocks.

\begin{figure}[h]
    \includegraphics[width=\columnwidth]{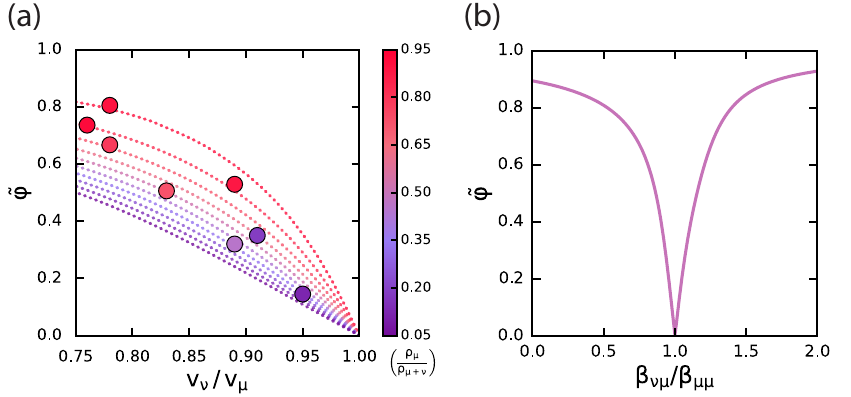}
    \caption{Demixing routes.
    (a) Self-propulsion speed difference. Variation of the demixing order parameter $\tilde{\varphi}$ with the speed ratio $v_\nu/v_\mu$. The total density is $\rho_{\mu + \nu} = 2.3 \times 10^{-3} \mu \rm m^{-2}$. The color codes for the density ratio of the mixtures.
    (b) Non-reciprocal repulsive interactions. Variations of $\tilde\varphi$ with the ratio of the repulsion strength $\beta_{\nu\mu}/\beta_{\mu\mu}$.}%
\label{Fig:Routes}
\end{figure}

\paragraph*{Spontaneous demixing routes.—}
The amount of demixing achieved in colloidal flocks is directly linked to the difference in self-propulsion speed between species (see SM Eq.~\eqref{Eq:RadialFractionSM}) and can be quantified by the order parameter:
 \begin{equation}
     \tilde{\varphi} = \left\langle  \left| \frac{\rho_\mu'(r) - \rho_\nu'(r)}{\rho_\mu'(r) + \rho_\nu'(r)} \right| \right\rangle_r,
          \label{Eq:OrderParameter}
 \end{equation}
where $\rho_\mu'(r)$ = $\rho_\mu(r) / \left\langle \rho_\mu(r) \right\rangle_r$.
$\tilde\varphi$ vanishes when both species occupy space homogeneously, and takes the value~$1$ when they are perfectly sorted.
Figure~\ref{Fig:Routes}(a) shows that $\tilde\varphi$ is predicted to increase from $0$ to $0.7$ when the speed ratio $v_\nu/v_\mu$ decreases from $1$ to $0.75$, in a 50:50 mixture at $\rho_{\mu + \nu} = 2.3\times10^{-3}\,\mu \rm m^{-2}$.
Strong demixing can be obtained for relatively small speed disparities, and the greater the speed disparity the stronger the demixing.
%
% We confirm this prediction by performing additional experiments.
% We realize binary colloidal flocks of varying mixture ratios for which we obtain slightly different speed ratios.

\textcolor{black}{
We confirm this prediction by studying binary colloidal flocks of varying mixture ratios.
In these additional experiments, we observe slight differences of the mean particle speeds (see SM Section~\ref{Sec:Experiments}).
While we do not explain these variations, we turn them into our advantage to test the effect of speed disparity on demixing.
}
As shown in Fig.~\ref{Fig:Routes}(a), the demixing in all flocks is in good agreement with the theory, confirming the accuracy of their hydrodynamic description.

Since self-propulsion speed differences induce the demixing of binary colloidal flocks, a natural question arises: Could non-reciprocal interactions alone drive the demixing of binary polar liquids?
The inspection of the structure of Eq.~\eqref{Eq:HydroCoefficients} helps answer this question.
In the case of colloidal rollers, non-reciprocity is reduced to two equal sets of actions: (i) the ones caused by species $\mu$ on {\em both} species $\mu$ and $\nu$, and (ii) the ones caused by species $\nu$ on {\em both} species $\mu$ and $\nu$.
As a consequence, the total alignment and the total repulsion acting upon both species are identical.
The microscopic non-reciprocity hence fades at the macroscopic scale, in a way similar to what has recently been reported for scalar active matter mixtures~\cite{dinelli2022nonreciprocity}.

We close this article by going beyond the current material constraints, and envision a generic polar liquid that does not adhere to the structure denoted in Eq.~\eqref{Eq:HydroCoefficients}, which opens a new route to achieve demixing. 
To this end, we consider a limiting case where the species $\mu$ experiences only homospecific interactions: $\alpha_{\mu\nu}$ = $\beta_{\mu\nu} = 0$.
Species $\mu$ therefore flocks as in a single-species system, forming a vortex of radially increasing density.
We further specify the behavior of species $\nu$ such that it evolves in this polar liquid background while only exhibiting heterospecific interactions ($\alpha_{\nu\nu} = 0$, $\beta_{\nu\nu} = 0$).
These interactions rules are summarized by:
\begin{equation}
\begin{gathered}
\underline{\underline{\mathbf{\alpha}}} = \alpha_{\mu \mu}
\begin{pmatrix}
1 & 0 \\
1 & 0 
\end{pmatrix},\;\;\;\;\;\;\;\;
\underline{\underline{\mathbf{\beta}}} = \beta_{\mu \mu}
\begin{pmatrix}
1 & 0 \\
\beta_{\nu \mu}/\beta_{\mu \mu} & 0 
\end{pmatrix}.
\end{gathered}
\label{Eq:HydroCoefficientsBis}
\end{equation}
The main control parameter in this simplified dynamics is $\beta_{\nu\mu}/\beta_{\mu\mu}$, which expresses the relative strength of repulsive interactions on species $\nu$ and $\mu$.
Simplifying Eqs.~\eqref{Eq:massBinary} and~\eqref{Eq:momentumBinary} with the relations of Eq.~\eqref{Eq:HydroCoefficientsBis} allows to obtain the particle density profiles varying $\beta_{\nu\mu}/\beta_{\mu\mu}$, see SM Section~\ref{Sec:HydroTheory}.
We find that such binary mixtures do spontaneously demix for identical active speeds $v_\mu = v_\nu$, see Fig.~\ref{Fig:Routes}(b).
For $\beta_{\nu\mu}/\beta_{\mu\mu} = 1$, no demixing occurs and $\tilde{\varphi} = 0$.
When $\beta_{\nu\mu}/\beta_{\mu\mu} < 1$, species $\nu$ tends to accumulate outwards, as its particles experience less repulsion at any given location than particles of species $\mu$.
Conversely, species $\nu$ accumulates inwards when $\beta_{\nu\mu}/\beta_{\mu\mu} > 1$. 
This behavior demonstrates how non-reciprocal interactions can exclusively lead to the demixing of binary polar liquids.
In general, both active speed differences and non-reciprocity will therefore contribute to setting the inner structure of binary polar liquids.

\paragraph*{Conclusion.—}

In this Letter, we have uncovered two mechanisms leading to the spontaneous demixing of active polar liquid vortices.
Both self-propulsion heterogeneity and non-reciprocity of binary interactions can drive the partial segregation of binary flocks.
The former is the principal driver of the demixing of binary colloidal flocks assembled from Quincke rollers of different sizes.
Accounting for non-reciprocal pairwise interactions between colloidal rollers allows for quantitatively capturing the binary flocks' spatial structuring.
Beyond its fundamental significance, our work could inspire the design of active sorting platforms which rely on active constituents' mobility and interactions rather than global mechanical motion.

\begin{acknowledgments}
We thank M. Le Blay for insightful discussions, and  M. Lettinga and V. Krabbenborg for their contributions at an early stage of this research.
\end{acknowledgments}

\end{document}